# Experimental Study of the Cloud Architecture Selection for Effective Big Data Processing

Evgeny Nikulchev

Moscow Technological Institute,
National Research University – Higher School of Economics
Moscow, Russia

Dmitry Biryukov

Moscow State Technical University of Radio Engineering, Electronics and Automatics
Moscow, Russia

Evgeniy Pluzhnik

Moscow Technological Institute
Moscow, Russia

Oleg Lukyanchikov

Moscow Technological Institute,
Moscow State Technical University of Radio Engineering, Electronics and Automatics
Moscow, Russia

Simon Payain

Moscow Technological Institute
Moscow, Russia

*Abstract*—**Big data dictate their requirements to the hardware and software. Simple migration to the cloud data processing, while solving the problem of increasing computational capabilities, however creates some issues: the need to ensure the safety, the need to control the quality during data transmission, the need to optimize requests. Computational cloud does not simply provide scalable resources but also requires network infrastructure, unknown routes and the number of user requests. In addition, during functioning situation can occur, in which you need to change the architecture of the application — part of the data needs to be placed in a private cloud, part in a public cloud, part stays on the client.**

## I. INTRODUCTION

Modern applications operate on large volumes of data that may reside in different stores. Cloud computing and cloud data storage are rapidly evolving, which gives advantages in performance due to parallel computing, the use of virtualization technology, scaling of computing resources and providing access to data via a web interface. Therefore, the actual task is to migrate existing systems and databases (DB) to the cloud.

Currently many developers and users are concerned about full advantage of cloud services. However, it is hard to tell in advance if a certain feature would be effective. Quite often new application features change the data structure. Furthermore, sometimes even a small modification can attract a large number of users, requiring database structure optimization to handle these large amounts of data. Currently the Big Data causes many problems for developers, since classic theories of query optimization only consider structure distribution and do not take inquiries depth into account. A new optimization parameter is introduced, called the data capacity with corresponding access and data transfer time. In this case normalized structures don't have to be optimal, based on the transfer time. This requires novel development techniques for data analysis, given the constantly growing amount of data and changing data structures. However, so far migration of existing systems to the cloud only creates problems. Security issues of access to data and guarantee of service can be solved by using a hybrid cloud - part of the data (processing of queries that require large computational resources and not confidential data) is placed in the public cloud services, and the remaining data — in the private cloud or local network infrastructure [1]. However, in this case specialized design principles of cloud systems are yet to be developed. In theoretical terms this problem was considered in [2–4]. There are few solutions for specific applications [5–7].

Complexity of construction techniques for distributed databases in a hybrid cloud is that it is impossible to estimate the parameters of algorithms and query performance: in each case the acquired amount of cloud resources, such as virtual machines, is different; routes and characteristics of communication channels are unknown. Optimization of the volume and types of resources are another important task. Therefore, at present, in the absence of developed general principles and methods, experiments represent the only way to study the effectiveness of design decisions for this research field.

There was created an experimental stand (ES) that simulates the work with hybrid storage. Stand itself and some experimental results obtained with it are described in [8, 9], see figure 1. Employed software VMWare vCloud allows you to organize at all levels. VMware ESXi is used on two servers to create a cloud in the ES. Management system VCenter and application VMware vCloud Director are deployed. In ES there are more than 15 physical Cisco 29 switches and routers Series 26 and Series 28, as well as virtual switches Nexus. System based on ES allows to simulate routes of access to the data, to converge and diverge channels (can be done dynamically).

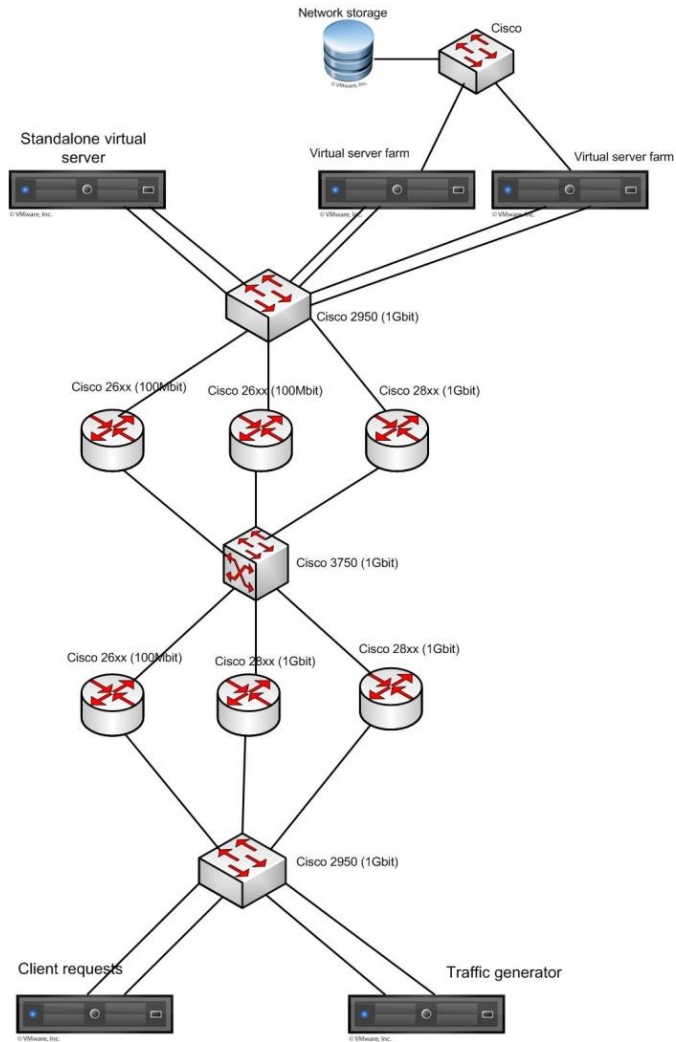

Fig. 1.   Experimental installation

## II.   EXPERIMENTS

For designed experimental setup, that emulates a hybrid cloud, experiment was conducted. Two types of database partitioning between public and private parts were examined. This is an important task for big data processing — in the process of operation and development of applications using the database with large volumes, including structured information, it is often required to transfer part of the data in the public cloud with a large number of resources, while not violating security requirements (i.e., leaving a portion of the data in the private cloud).

The aim of this experiment is to determine the efficiency of separation of the DB into 2 parts: public and private. For the experiment 3 compute nodes were prepared, the overall structure of which is shown in figure 2:

- Public DBMS server, which is more powerful (dual-core processor, 4 GB of RAM).

- Private DBMS server, which is less powerful (single-core processor, 2 GB of RAM).

- A client that makes requests to the published server using specialized software.

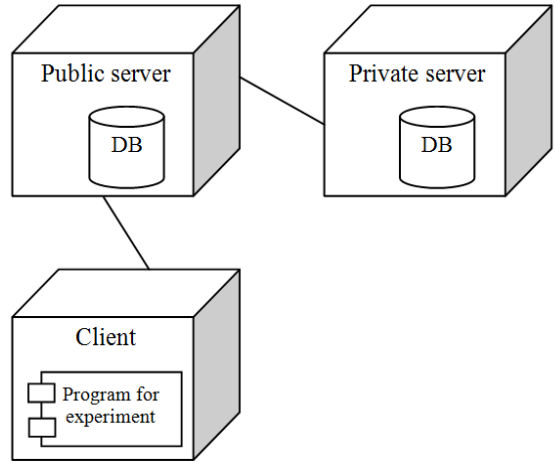

Fig. 2.   The structure of the experiment

For the experiment there was used part of the database included in the educational process at the University (figure 3).

Table "Students" contains the following information about students:

- "id_stud" - unique number of the student, it is the primary key;

- "fio" - surname, name and second name of the student;

- "birthday" - the date of student's birth;

- "agv_score" - the average score;

- "group" - the group number, which includes the student; is an external key of the table "Groups".

Table "Groups" associated one-to-many with table "Students" contains the following information about groups:

- "id_group" - unique group number;

- "name" - the name of the group.

Table "Lections" associated many-to-many with table "Students" via table "visit", contains following information about lectures:

- "id_lection" - unique number of lectures;

- "subject" - the number of the held object;

- "room" - the cabinet number (the audience);

- "theme" - is the theme of the lecture;

- "date" - the date of the lecture.

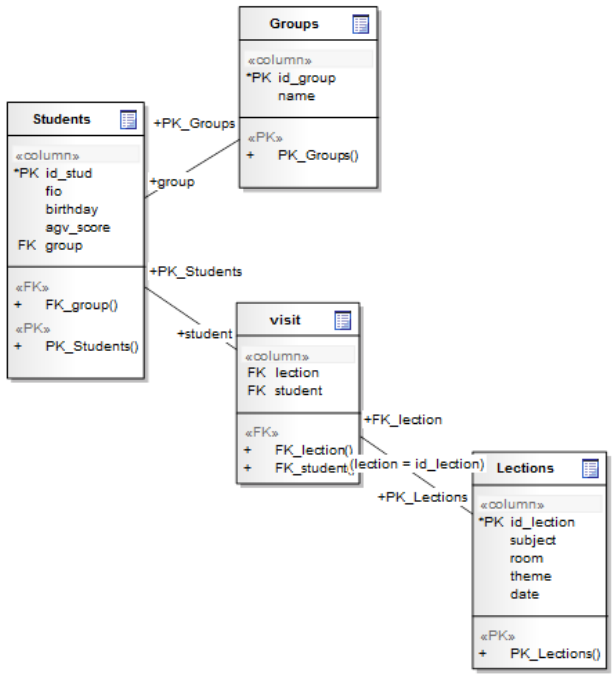

Fig. 3.   The database scheme

Table "visit" includes the attendance of students.

- "id_student" - the number of student. Is external to the key of the table "Students";

- "id_lection" – the number of the lecture. Is external key to the table "Lections".

For this experiment the database was filled with random data:

Table "Students" - 3 000 000 records;

Table "Groups" - 100 records;

Table "Lections" - 100 000 records;

Table "visit" - 100 000 records.

The results of the query fetching all data from table "students" (adding data from tables associated one-to-many and many-to-many) were compared to test the effectiveness of the separation of the database into public and private parts.

In the first case, the entire database was on a public server, to retrieve the data following query was used:

*select * from students.students*

*left join students.groups on students."group" = groups.id_group*

*left join (select * from students.visit*

*left join students.lections on visit.id_lection = lections.id_lection) t1*

*on students.id_stud = t1.id_student.*

In the second case the part that relates many-to-many with table "Students" was placed on a private server (figure 4). Data was obtained by the function PostgreSQL dblink, which allows to perform the query to another DBMS. Request in the second case:

*select * from students.students*
*left join students.groups on students."group" = groups.id_group*
*left join (select * from dblink('hostaddr=xxx.xxx.xxx.xxx port=xxxx dbname=… user=… password=…', 'select id_student,lections.id_lection,id_subject,date,room,theme from students.visit*

*left join students.lections on visit.id_lection = lections.id_lection') as t(id_student INTEGER,id_lection INTEGER,id_subject INTEGER,date DATE,room INTEGER,theme TEXT)) as t1*
*on students.id_stud = t1.id_student*

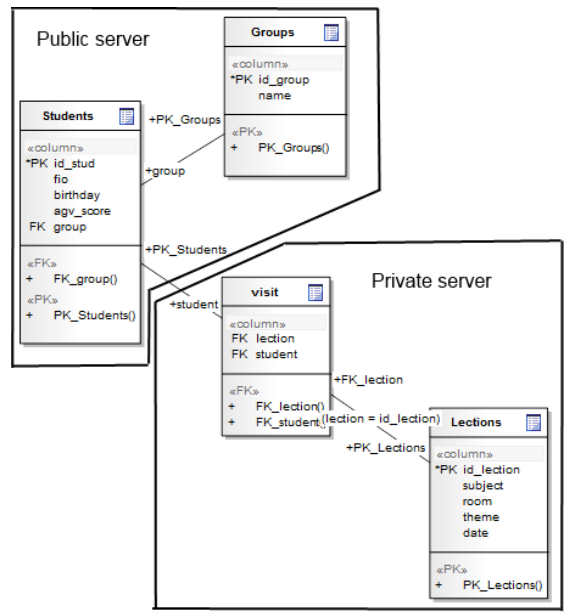

Fig. 4.   Split the database between the public and private servers

Using such functions as dblink (which allows to perform queries to another DBMS), makes it possible to bring part of the database to another DBMS with changes to the query without need of modifying client application.

Effectiveness evaluation was performed using specialized software on the client, which emulates operation of the client application. To do this, through a small random amount of time the application has been generating queries, getting data about several students.

In the first case, when the entire database is on a public server, for 10 minutes 130 requests were performed. Time efficiency is presented in figure 5 (on the x-axis is the number of the query, on the y-axis is the query execution time in seconds).

In the second case, when the database is divided between public and private servers, for 10 minutes 126 requests were performed. Time efficiency is presented in figure 6 (on the x-axis is the number of the query, on the y-axis is the query execution time in seconds).

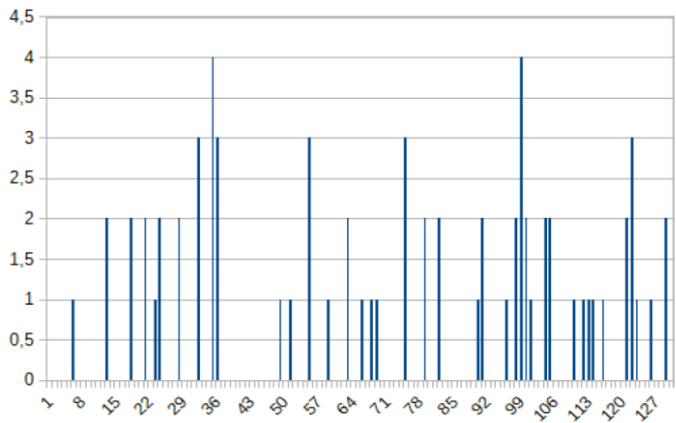

Fig. 5. Time efficiency when performing queries on the same public server

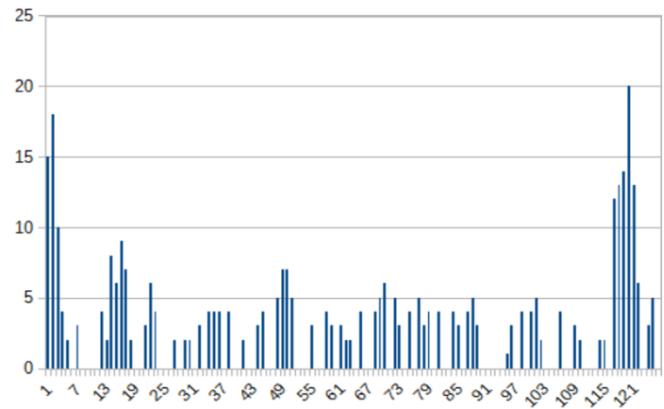

Fig. 1. Time efficiency when performing queries to public and private servers

## III. DISCUSSION

Main feature of the application is the intermediate layer that implements connection of user requests to the location of distributed data. Presence of unknown destination switching when using public cloud and mobile client makes it impossible to estimate the time of the algorithms. Here is why it's advisable to use software technology to control all stages of the system. However, hybrid infrastructure has many positive aspects of cloud computing: scalability, virtualization and also (due to the distribution of data) safety and security of data; designing information systems in the cloud has following problems:

- Impossible to assess the execution of individual queries and stream query;

- No general principles for designed systems with large amounts of data (Big data);

- Considerable amount of educational and scientific data is semistructured (XML);

- No technology for migration of database to the cloud – need to rewrite the code when moving to hybrid cloud;

- No commonly accepted principles for virtualization management allocation of resources in the cloud;

- The limitations associated with using an obsolete protocol (TCP).

The task was to account these features and develop the technology that creates applications in a hybrid cloud.

Within these limitations, the principles of development that provide guaranteed quality and functioning of the application were developed:

*1) The system design should be based on a preliminary study on the simulation and experimental models.*

*2) It is necessary to control the main parameters of the infrastructure.*

*3) The use of object-oriented technology modifications of database design.*

*4) Technology should provide the flexibility of system structure, data volume, number of requests.*

The result shows that in the second case, when the database is divided, the average query execution time is much higher than with solid database. It means that the separation of the database on the public and private parts adversely affects the performance of the system, and only has advantages from the viewpoint of safety.

Distributed data complicates the development of software, making it difficult and time-consuming to use common programming techniques. Despite development of technologies such as .Net and Qt, developers eventually have to operate SQL queries and clearly prescribe access to the distributed data. In the context of widespread object-oriented development methodology and application systems, with relational DBMS having dominant position in the market, advisable solution is to use intermediate software that provides necessary object-oriented interface to the data stored under control of a relational DBMS. To communicate with developed relational data objects there was used Object Relational Mapping (ORM) [10]. The essence of this technology is in accordance of programming entity to relational database object: each field of a table is assigned to a class attribute of an object.

The basic steps are the following:

*1) Determination of basic structure of physically distributed in the hybrid cloud data.*

*2) Development of relational database structure.*

*3) Development of methods for data processing based on physical location of the data.*

*4) Creating classes of objects, including data and methods for their treatment.*

*5) Modification of the structure as a result of experimental study on the simulation bench.*

*6) Changing methods of processing inheritance*

Depending on the task, system can be restructured to increase the speed of the most common queries, or to perform the most demanding requests in the public cloud.

## IV. CONCLUSION

Thus, based on experiments with cloud infrastructure and different database decompositions, following features were observed:

-The need to use specialized tools to control feedback on routing and protocol levels.

-It is required, starting with development phase of the software application, to include the possibility of changing the database structure.